\documentclass[aps,prb,reprint,superscriptaddress]{revtex4-1}

\usepackage{amsmath}    % need for subequations
\usepackage{graphicx}   % need for figures
\usepackage{color}      % use if color is used in text
\usepackage{subfigure}  % use for side-by-side figures
\usepackage{todonotes}
\usepackage{nicefrac}
\usepackage[utf8]{inputenc}

\usepackage{hyperref}   % use for hypertext links, including those to external documents and URLs

\begin{document}

\title{Evidence for phase formation in potassium intercalated 1,2;8,9-di\-benzo\-penta\-cene}

\author{Friedrich Roth}
\author{Andreas K\"onig}
\author{Benjamin Mahns}
\author{Bernd B\"uchner}
\author{Martin Knupfer}
\affiliation{IFW Dresden, P.O. Box 270116, D-01171 Dresden, Germany}
\date{\today}

\begin{abstract}
We have prepared potassium intercalated 1,2;8,9-di\-benzo\-penta\-cene films under vacuum conditions. The evolution of the electronic excitation spectra upon potassium addition as measured using electron energy-loss spectroscopy clearly indicate the formation of particular 
doped phases with compositions K$_x$di\-benzo\-penta\-cene ($x$ = 1,2,3). Moreover, the stability of these phases as a function of temperature 
has been explored. Finally, the electronic excitation spectra also give insight into the electronic ground state of the potassium 
doped 1,2;8,9-di\-benzo\-penta\-cene films.
\end{abstract}

\maketitle

\section{Introduction}
Molecular crystals---built from $\pi$ conjugated mole\-cules---are in the focus of research for a number of reasons. Within this class of
materials, almost every ground state can be realized at will, spanning from insulators to semiconductors, metals, superconductors or magnets.
Due to their relatively open crystal structure their electronic properties can be easily tuned by the addition of electron acceptors and donors.
In some cases, this resulted in intriguing and unexpected physical properties. For instance, carbon based superconductors have a long history
dating back to 1965, when superconductivity was found in alkali-metal doped graphite, with transition temperatures of
$T_c$\,$<$\,1~\cite{Hannay1965}. Recently, the $T_c$ was increased up to 11.5\,K for calcium intercalated graphite~\cite{Emery2005}.

\par

However, more than 25 years later the discovery of a superconducting phase in the alkali metal doped fullerides \cite{Hebard1991,Gunnarsson2004}
represented a breakthrough in the field of superconductivity and attracted a lot of attention, also because of rather high transition
temperatures up to 40\,K~\cite{Tanigaki1991,Ganin2008,Ganin2010,Palstra1995}. In this context, further interesting phenomena were observed in
alkali metal doped molecular materials such as the observation of an insulator-metal-insulator transition in alkali doped phthalocyanines
\cite{Craciun2006}, a transition from a Luttinger to a Fermi liquid in potassium doped carbon nanotubes \cite{Rauf2004}, or the formation of a
Mott state in potassium intercalated pentacene~\cite{Craciun2009}.

\par

In the case of organic superconductors, transition temperatures similar to those of the fullerides could not be observed in other molecular
crystals until 2010, when superconductivity has been reported for another alkali metal doped molecular solid, K-picene, with a $T_c$ up to
18\,K~\cite{Mitsuhashi2010}. Furthermore, after this discovery superconductivity was also reported in other alkali metal intercalated polycyclic
aromatic hydrocarbons, such as K-phenanthrene ($T_c$ = 5\,K) \cite{Wang2011,Andres2011}, K-coronene ($T_c$ = 15\,K)\cite{Kubozono2011} and
K-(1,2;8,9-di\-benzo\-penta\-cene) ($T_c$ = 33\,K)~\cite{Xue2011}. Especially in the latter case, the $T_c$ is higher than in any other organic
superconductor besides the alkali-metal doped fullerides. Now, a thorough investigation of the physical properties of 1,2;8,9-di\-benzo\-penta\-cene
in the undoped as well as in the doped state is required in order to develop an understanding of the superconducting and normal state
properties.

\par

Apart from the introduction of charge carriers, the addition of potassium to di\-benzo\-penta\-cene can also lead to stable phases with
particular stoichiometries. A very important prerequisite for detailed studies as well as the understanding of physical properties is the
knowledge about such phases and their existence and stability regions. For instance, the physical properties and the conclusive analysis of
experimental data of alkali metal fullerenes have been demonstrated to be strongly dependent on the existing phases and their characterization~\cite{Gunnarsson2004,Ganin2008,Ganin2010,Takabayashi2009,Benning1992,Stepniak1993,Weaver1992,Poirier1995,Knupfer1994,Rosseinsky1995,Pichler1994,Kuzmany1995}.

\par

In this contribution we report on an investigation of the structural and electronic properties of potassium doped di\-benzo\-penta\-cene using
electron energy-loss spectroscopy (EELS) in transmission. EELS studies of other undoped and doped molecular materials in the past have provided
useful insight into their electronic properties~\cite{Schuster2007,Roth2010_2,Knupfer1999_2}. We discuss the changes that are induced in the
electronic excitation spectrum as a function of doping, and we provide clear evidence for the formation of three doped phases with
K$_1$C$_{30}$H$_{18}$, K$_2$C$_{30}$H$_{18}$ and K$_3$C$_{30}$H$_{18}$ composition. Moreover, temperature dependent investigations also allowed
first insight into the stability regions of those phases, and the electronic excitation spectra suggest insulating as well as metallic ground
states.

\section{Experimental}

\begin{figure}[ht]
\centering
\includegraphics[width=0.7\linewidth]{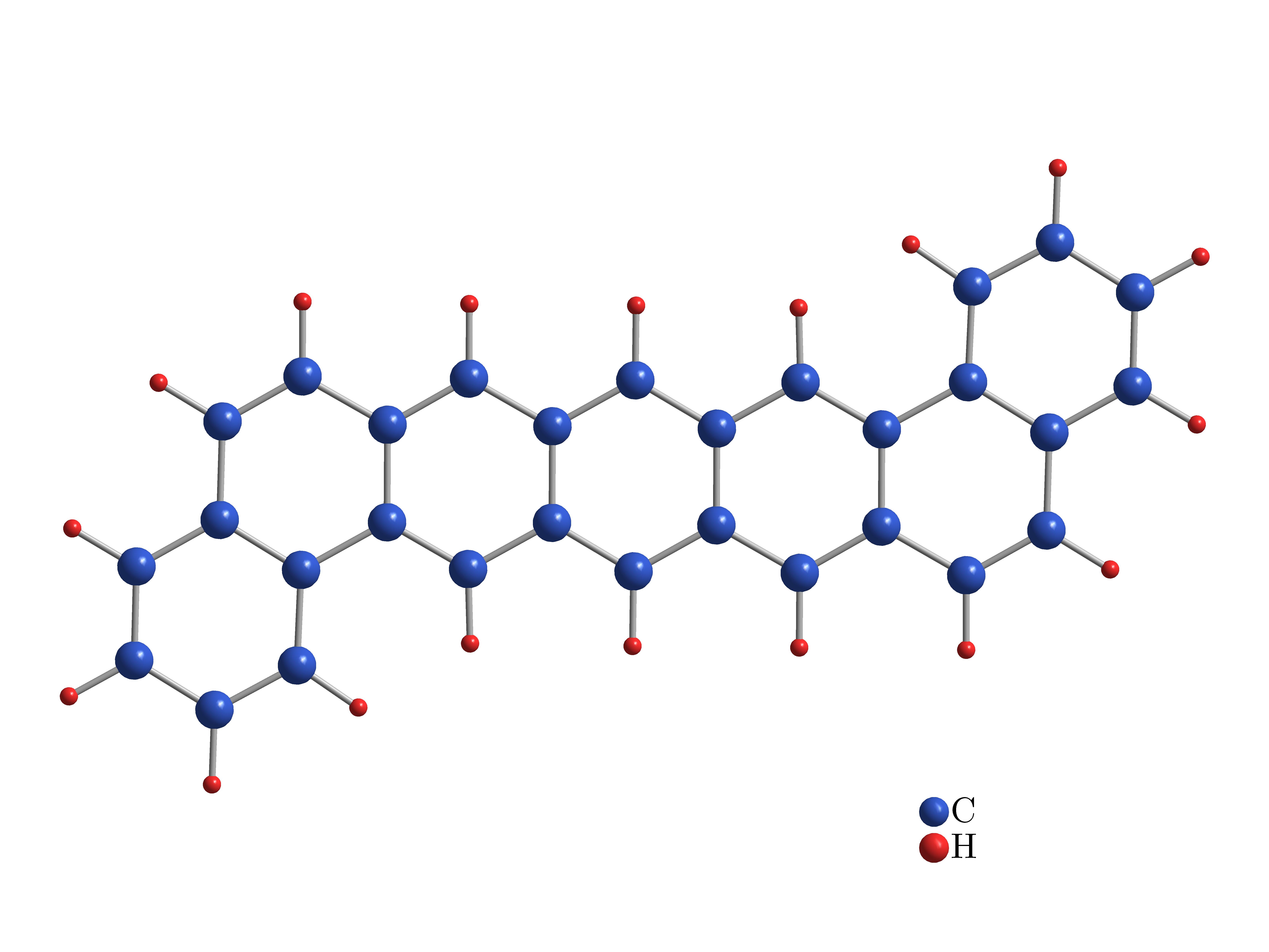}
\caption{Schematic representation of the molecular structure of 1,2;8,9-di\-benzo\-penta\-cene (C$_{30}$H$_{18}$).} \label{fig:1}
\end{figure}

1,2;8,9-di\-benzo\-penta\-cene (C$_{30}$H$_{18}$) is a molecule formed by seven benzene rings as depicted in Fig.\,\ref{fig:1}. It looks like a pentacene
molecule with one snapped off benzene ring on both ends. Up to now, no details of the crystal structure are published.

\noindent Thin films of di\-benzo\-penta\-cene (BGB Analytik AG, 4461 Boeckten, Switzerland) were prepared by thermal evaporation under high vacuum conditions (base pressure lower than 10$^{-8}$\,mbar) onto single crystalline KBr substrates 
kept at room temperature with a deposition rate of 0.3\,nm/min and an evaporation temperature of about 530\,K. The film thickness was about
100\,nm. These di\-benzo\-penta\-cene films were floated off in destilled water, mounted onto standard electron microscopy grids and
transferred into the spectrometer. Prior to the EELS measurements the films were characterized \emph{in~situ} using electron diffraction. The
diffraction spectra show no significant pronounced texture which leads to the conclusion that our films are essentially polycrystalline.

\par

All measurements were carried out using the 172\,keV spectrometer described in detail elsewhere~\cite{Fink1989}. We note that at this high
primary beam energy only singlet excitations are possible. The energy and momentum resolution were chosen to be 85\,meV and 0.03\,\AA$^{-1}$,
respectively. We have measured the loss function Im[-1/$\epsilon(\textbf{q},\omega)$] for a momentum transfer $\textbf{q}$ parallel to the film
surface, which probes the electronic excitations of the films [$\epsilon(\textbf{q},\omega)$ is the dielectric function]. In addition, the
C\,$1s$ and K\,$2p$ core level excitations were measured with an energy resolution of about 200\,meV and a momentum resolution of 0.03\,\AA. In
order to obtain a direction independent core level excitation information, we have determined the core level data for three different momentum
directions such that the sum of these spectra represent an averaged polycrystalline sample~\cite{Egerton1996}. The core excitation spectra have been corrected for
a linear background, which has been determined by a linear fit of the data 10\,eV below the excitation threshold. Since molecular crystals often are damaged by fast electrons, we repeatedly checked our samples for any sign of degradation. In particular, degradation was followed by watching an increasing amorphous-like background in the electron diffraction spectra and a broadening of the spectral features in the loss function. It turned out that under our measurement conditions the spectra remained unchanged for about 14\,h at 20\,K and 8\,h at room temperature. Samples that showed any signature of degradation were not considered further but replaced by newly prepared thin films. The 
results from the different films have been shown to be reproducible.

Potassium was added in several steps by evaporation from a commercial SAES (SAES GETTERS S.p.A., Italy) getter source under ultra-high vacuum
conditions (base pressure lower than 10$^{-10}$\,mbar). In each doping step, the sample was exposed to potassium for several minutes, the
current through the SAES getter source was 6\,A and the distance to the sample was about 30\,mm. Further details of the doping procedure are
discussed below.

\section{Results and discussion}
\par

\begin{figure}[ht]
\centering
\includegraphics[width=0.7\linewidth]{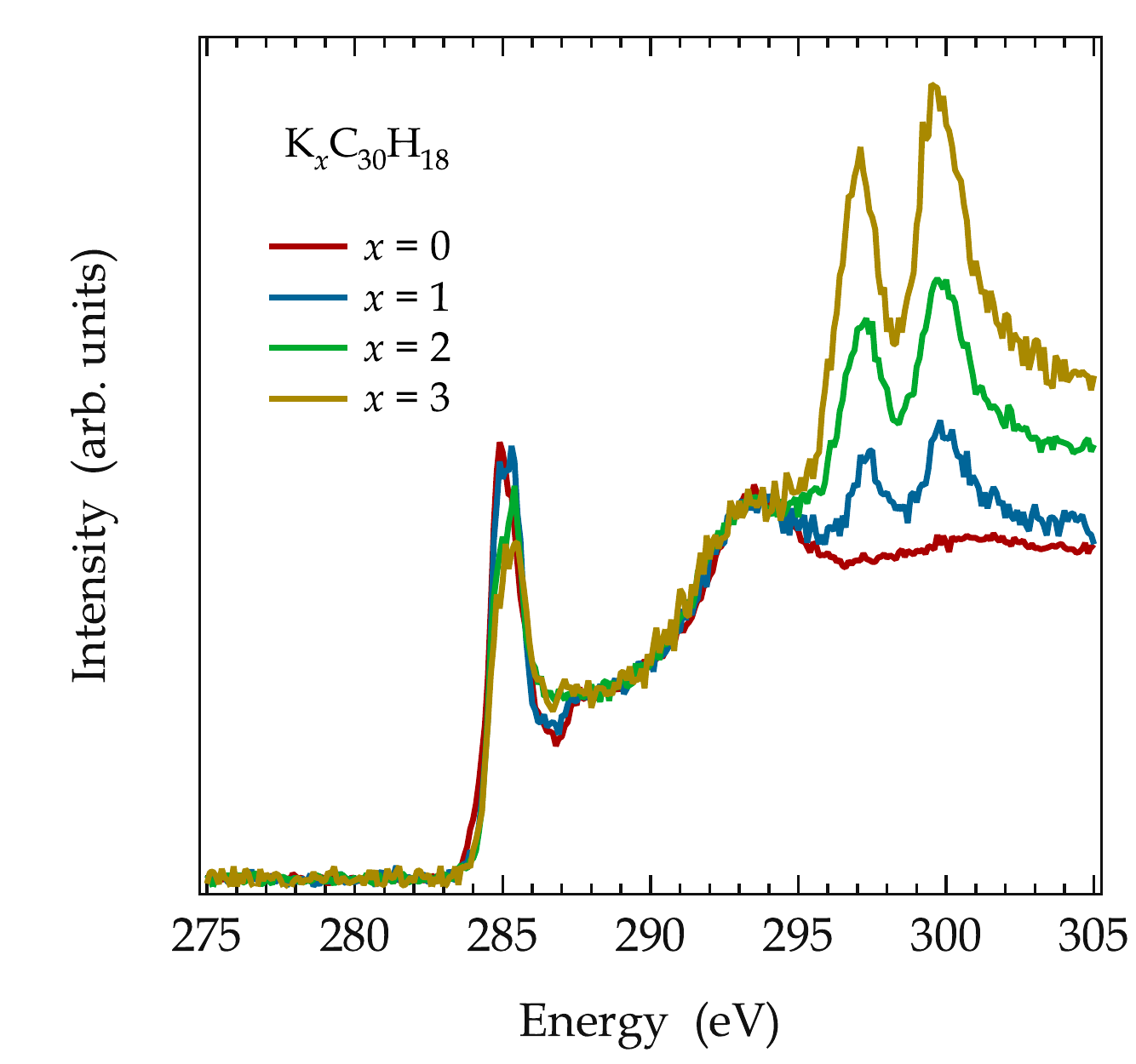}
\caption{ C\,$1s$ and K\,$2p$ core level excitations of K$_x$(1,2;8,9-di\-benzo\-penta\-cene) for $x$ = 0,1,2,3.} \label{fig:2}
\end{figure}

The amount of potassium in our doped di\-benzo\-penta\-cene films was determined using core level excitation spectra. In Fig.\,\ref{fig:2} we show C\,$1s$
and K\,$2p$ core level excitations of undoped and potassium doped di\-benzo\-penta\-cene. These data can be used to analyze the doping induced
changes of the potassium doped films.  Moreover, the C\,$1s$ excitations represent transitions into empty C\,$2p$-derived levels, and thus allow
to probe the projected unoccupied electronic density of states of carbon-based materials~\cite{Roth2008,Roth2010,Knupfer1999,Knupfer1995}. All
spectra were normalized at the step-like structure in the region between 291\,eV and 293\,eV, i.\,e. to the $\sigma^*$ derived intensity, which
is proportional to the number of carbon atoms. For the undoped case (red line), we can clearly identify a sharp and strong feature in the range
between 283 - 286\,eV, which  can be assigned to transitions into $\pi^*$ states representing the unoccupied electronic states. The step-like
structure above 291\,eV corresponds to the onset of transitions into $\sigma^*$-derived unoccupied levels.

\par

By doping the sample with potassium the spectrum is still dominated by a sharp excitation feature right after the excitation onset at 283\,eV and, in addition, by K\,$2p$ core excitations, which can be observed at 297.2\,eV and 299.8\,eV, and which can be seen as a first evidence of the successful doping of the sample. Importantly, a reduction of the spectral weight of the first C\,$1s$ excitation feature is observed in Fig.\,\ref{fig:2} upon doping, which can be seen as a further signal of succesfully doping because it represents the filling of the conduction band.

\par

The stoichiometry analysis can be substantiated by comparing the K\,$2p$ and C\,$1s$ core excitation intensities in comparison to other doped
molecular films with well known stoichiometry, such as K$_6$C$_{60}$~\cite{Knupfer2001}. Details of this procedure can be found in previous
publications~\cite{Flatz2007,Roth2008}. The results shown in the Fig.\,\ref{fig:2} indicate three different doping levels with K$_1$C$_{30}$H$_{18}$,
K$_2$C$_{30}$H$_{18}$ and K$_3$C$_{30}$H$_{18}$ composition, which are discussed in more detail in the following.

\begin{figure*}[ht]
\centering
\includegraphics[height=6cm]{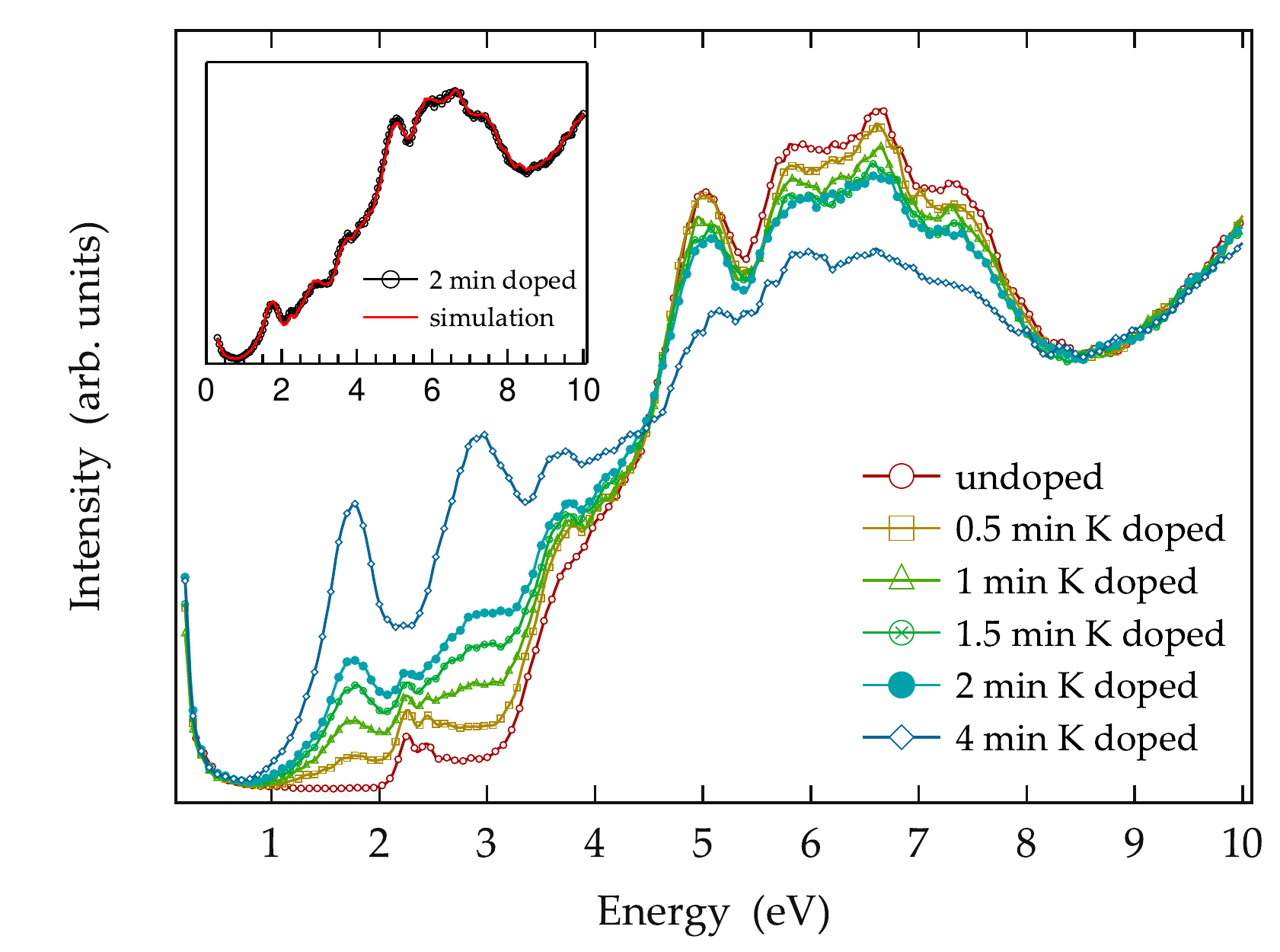}
\includegraphics[height=6cm]{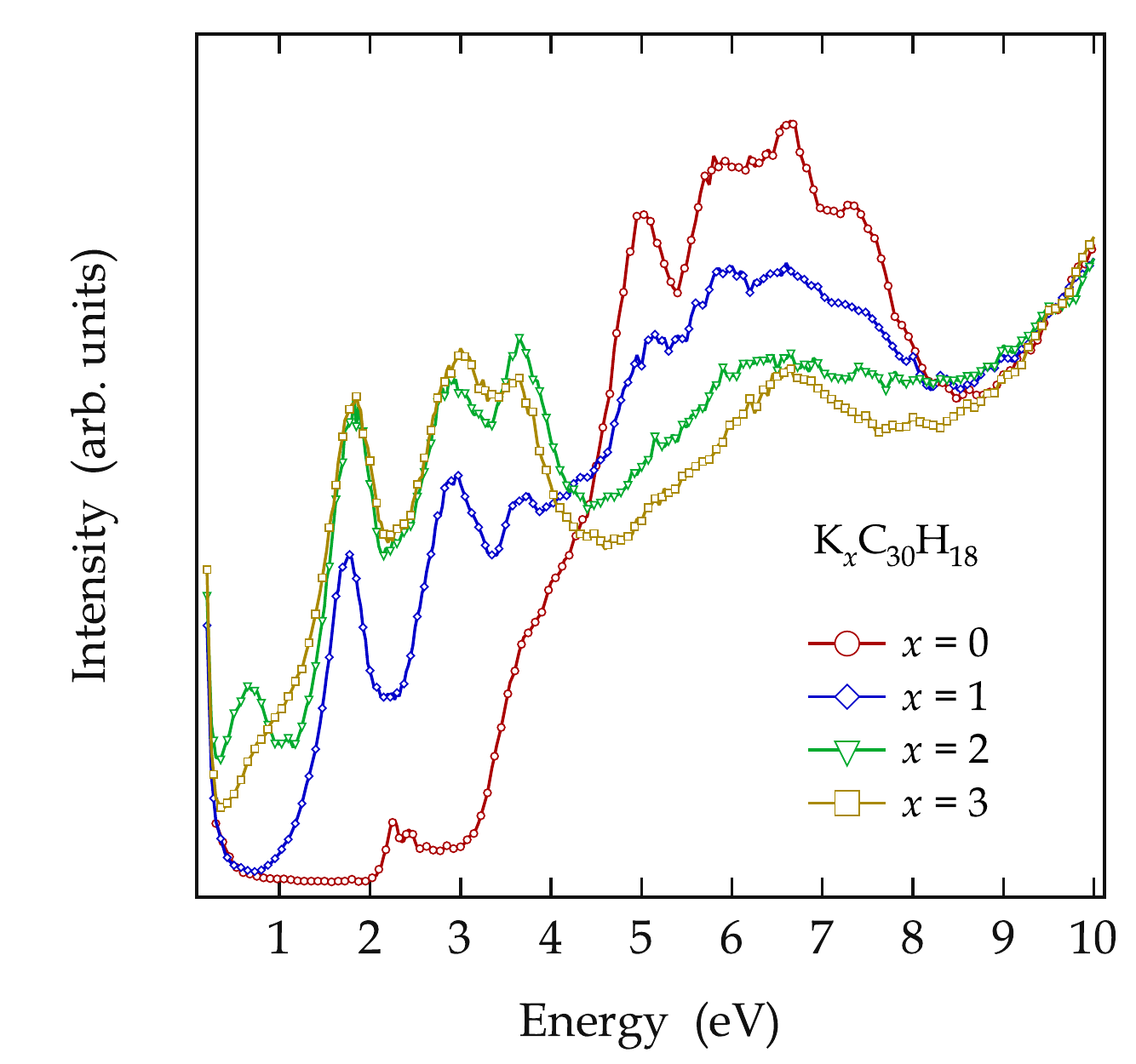}
\caption{Left panel: Evolution of the loss function of di\-benzo\-penta\-cene in the range of 0 - 10\,eV upon potassium doping measured with a momentum transfer of $\textbf{q}$ = 0.1\,\AA$^{-1}$. All spectra were normalized in the high energy region between 9 - 10\,eV. (K content increases from bottom (red open circles) to top (blue open diamonds)). The inset shows a simulation of the 2 min doped spectra as a superposition of 55\% of the undoped spectra and 45\% of the 4 min doped spectra. Right panel: Comparison of the loss function for different achieved doped phases.}
\label{fig:3}
\end{figure*}

Initial potassium addition to di\-benzo\-penta\-cene causes major changes in the electronic excitation spectrum as revealed in Fig.\,\ref{fig:3}
(left panel), where we show a comparison of the spectra in an energy range of 0-10\,eV measured using EELS for different doping steps. These
data are taken with a small momentum transfer $\textbf{q}$ of 0.1\,\AA$^{-1}$, which represents the optical limit. During potassium addition, the
di\-benzo\-penta\-cene films have been kept at room temperature.

\par

For undoped di\-benzo\-penta\-cene (red open circles), we can clearly identify maxima at about 5\,eV, 5.9\,eV, 6.6\,eV and 7.3\,eV as well as a broad
shoulder at about 3.75\,eV, which are due to excitations between the occupied and unoccupied electronic levels. Zooming into the energy region
around the excitation onset reveals an optical gap of 2\,eV. This onset also represents a lower limit for the band gap (or transport energy gap)
of solid di\-benzo\-penta\-cene. The excitation onset is followed by three rather weak electronic excitations at about 2.28\,eV, 2.43\,eV, and
2.62\,eV. The main features of our spectrum are in good agreement with previous optical absorption measurements in
solution~\cite{Clar1943,Clar1962,Perkampus1963}. In general, the lowest electronic excitations in organic molecular solids usually are excitons,
i.\,e. bound electron-hole pairs~\cite{Pope1999,Lof1992,Hill2000}. We assume that this is also true for di\-benzo\-penta\-cene, a detailed
analysis however of e.\,g. the exciton binding energy requires the knowledge of the so-called transport energy gap, which to our knowledge has not
been determined yet.

\par

Fig.\,\ref{fig:3} (left panel) reveals that upon initial doping, the spectral features become broader. The low energy structures representative of
undoped di\-benzo\-penta\-cene decrease in intensity while three new peaks become visible at 1.76\,eV, 2.93\,eV and 3.7\,eV. The latter steadily
increase with doping until a particular doping level  (labelled with 4 min K doped in Fig.\,\ref{fig:3}) is reached. Most importantly, all spectra in
the series as shown in Fig.\,\ref{fig:3} (left panel) can be simulated by a corresponding superposition of the spectra of undoped and 4 minutes doped
di\-benzo\-penta\-cene. This is demonstrated in the inset in Fig.\,\ref{fig:3}, where we show a comparison of the spectrum of a 2 min doped film and a
superposition of the two spectra of undoped and 4 min doped di\-benzo\-penta\-cene weighted by 0.55 and 0.45, respectively. In addition, further
potassium addition causes qualitative changes of the spectral shape, in particular the appearance of a new feature at at 0.65\,eV (see right
panel of Fig.\,\ref{fig:3}). Consequently, these two facts strongly indicate the formation of a potassium doped di\-benzo\-penta\-cene phase, and
our core level measurements signal that we reached a doping level of x = 1. Interestingly, this K$_1$di\-benzo\-penta\-cene is stable up to
100\,$^{\circ}$C, i.\,e. heating up the sample to this temperature did not induce visible changes in the valence band as well as in the core level
spectra. Going to 150\,$^{\circ}$C however resulted in a loss of potassium from the films as could be seen by a reduction of the excitation
feature at 1.76\,eV. In regard of the electronic ground state of the phase with K$_1$di\-benzo\-penta\-cene composition, the data in Fig.\,\ref{fig:3}
indicate an energy gap of about 0.9\,eV, i.\,e. this K$_1$-phase is insulating.

\par

Starting from the above discussed phase (K$_1$di\-benzo\-penta\-cene) and adding further potassium while the films are kept at room temperature leads
to a second doped phase, which is characterized by the additional peak in the electronic excitation spectrum at 0.65\,eV, and which according to
our core level analysis has a stoichiometry of K$_2$di\-benzo\-penta\-cene. Moreover, adding more potassium at this stage leads to the formation of a
potassium overlayer on our films as signalled by the appearance of the charge carrier plasmon excitation of metallic potassium at 3.75\,eV. This indicates that there is no further diffusion of potassium into a film with K$_2$di\-benzo\-penta\-cene
composition at room temperature. Intriguingly, keeping the films at room temperature in our ultra high vacuum chamber results in a loos of
potassium after several minutes as signalled by a loss the spectral weight around 0.65\,eV, i.\,e. potassium diffuses out of the film. Cooling a
K$_2$di\-benzo\-penta\-cene film down to 20\,K, however, allows to keep the composition stable for at least 15 hours. Again spectra with intermediate composition can be well simulated by the superposition of the spectra of K$_1$di\-benzo\-penta\-cene + K$_2$di\-benzo\-penta\-cene.

\par

To summarize at this point, we have prepared two potassium doped phases of di\-benzo\-penta\-cene, K$_1$di\-benzo\-penta\-cene and
K$_2$di\-benzo\-penta\-cene, which can be achieved by potassium addition at room temperature under UHV conditions and which are well distinguished by
corresponding features in the electronic excitation spectra. The temperature stabilty of these two phases is rather different since the former
can be heated to about 100\,$^{\circ}$C without noticeable changes while the latter is only stable at very low temperatures (20\,K).

However, the observation of superconductivity was reported for samples with even higher doping level, between x = 3 and 3.5. In order to realize
such high doping levels, we had to change our doping procedure. In detail, we heated up the films during potassium addition (which lasted 15
minutes) to temperatures of 60 - 80\,$^{\circ}$C, and we annealed the films after each doping step at 80\,$^{\circ}$C for about 15 minutes.
Finally, at least three doping and annealing cycles were necessary to identify clear changes in the electronic excitation spectra as revealed in
Fig.\,\ref{fig:3} (right panel). The feature typical for K$_2$di\-benzo\-penta\-cene at 0.65\,eV disappears and instead a shoulder centered at 0.85\,eV
shows up. Furthermore, there are slight changes in the double peak structure at 2.93\,eV and 3.7\,eV. Our measurements of the C\,$1s$ and
K\,$2p$ core level excitations display a doping level of x $\approx$ 3 (cf. Fig.\,\ref{fig:2}). The necessary annealing during and after potassium
addition clearly signals, that potassium diffusion into the di\-benzo\-penta\-cene films requires much higher activation energy at higher doping
levels. Interestingly, the observation of superconductivity in samples with about three potassium per di\-benzo\-penta\-cene molecule \cite{Xue2011}
was made after long time annealing of the doped samples which also indicates hindered potassium diffusion at room temperature.

Similar to what we have observed for the K$_1$di\-benzo\-penta\-cene phase, our films with a K$_3$di\-benzo\-penta\-cene composition also are stable at
100\,$^{\circ}$C for more than 15 hours. Further, higher temperatures (150\,$^{\circ}$C) again results in a loss of potassium from the films,
and we observe spectral changes towards the spectral shape as observed for K$_1$di\-benzo\-penta\-cene. Consequently, this infers a potassium binding
energy that is rather similar for these two potassium doped di\-benzo\-penta\-cene phases.

\par

Finally, the excitation spectra for the K$_2$- and the K$_3$-phase do not show clear evidence for an energy gap in contrast to what is observed for K$_1$dibenzopentacene. Also measurements with higher momentum transfer do not show the opening of a band gap. This could be taken as evidence for a metallic ground state of these phases, in agreement with the observation of a superconducting phase at higher doping levels \cite{Xue2011}. However, the presence of the elastic line in our spectra does not allow the determination of the true ground state, metallic or insulating. In general, the doping induced excitation at about 1.76\,eV could be interpreted as a transition from the now filled lowest unoccupied molecular orbital (LUMO) of dibenzopentacene to the LUMO+1 level. The lowest energy feature around 0.65\,eV then could be interpreted as a charge carrier plasmon of the metallic phases. On the other hand, recent photoemission investigations of potassium doped picene and coronene, two other recently reported superconductors, were not able to identify any metallic doped phase, which might be related to the importance of electron-electron correlation effects \cite{Mahns2012}. Given these facts, further work is required in order to unambiguously demonstrate well defined and well characterized metallic phases of doped hydrocarbon molecular solids.

\section{Summary}

\begin{figure}[ht]
\centering
\includegraphics[width=0.7\linewidth]{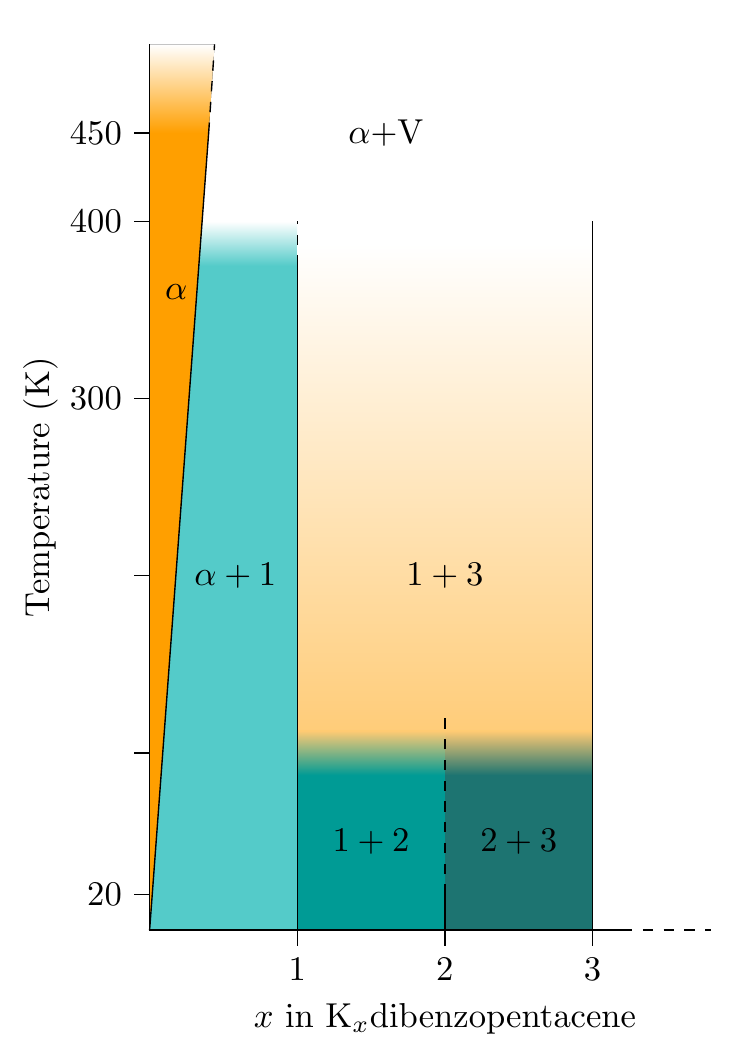}
\caption{Provisional phase diagram of K$_x$(1,2;8,9-di\-benzo\-penta\-cene) for $x$ up to 3. In mixed phase regions, the corresponding phases are
denoted by their x value, V stands for potassium vapor.} \label{fig:4}
\end{figure}

To conclude, we have investigated potassium doping of thin films of di\-benzo\-penta\-cene using electron energy-loss spectroscopy. The doping
induced changes in valence band as well as core level excitations, and the temperature dependence of the measured spectra clearly indicate the
formation of phases with K$_1$di\-benzo\-penta\-cene, K$_2$di\-benzo\-penta\-cene, and K$_3$di\-benzo\-penta\-cene composition. Our results thus provide insight
into the phase behavior of K$_x$di\-benzo\-penta\-cene, which we summarize in a provisional phase diagram as depicted in Fig.\,\ref{fig:4}. This schematic
diagram also includes a solid solution of potassium in the di\-benzo\-penta\-cene films denoted $\alpha$, which would form first upon doping. We note
that all our experiments have been carried out at a pressure of about 10$^{-10}$\,mbar, and that the temperatures as given in Fig.\,\ref{fig:4} do not
represent the atmospheric pressure behavior. Furthermore, film growth and doping by deposition from the vapor phase is characterized by the
interplay between thermodynamics and kinetics, which can be nicely seen by the annealing steps that are necessary to achieve a doping level of
K$_3$di\-benzo\-penta\-cene. Finally, our data suggest that K$_1$di\-benzopentacene has an insulating ground state with an energy gap of about 0.9\,eV,
while K$_2$di\-benzo\-penta\-cene and K$_3$di\-benzo\-penta\-cene might well be metallic, because we do not find signatures of an energy gap in the
electronic excitation spectra.

\par

\begin{acknowledgments}
 We thank R. Sch\"onfelder, R. H\"ubel and S. Leger for technical assistance. This work has been supported by the Deutsche
Forschungsgemeinschaft (grant number KN393/14).
\end{acknowledgments}

\end{document}